\documentclass[aps,pre,twocolumn,superscriptaddress,showpacs]{revtex4-1}

\usepackage{amssymb}
\usepackage{color}
\usepackage{graphicx}
\usepackage{amsmath}
\usepackage{times}
\usepackage{bm}

\begin{document}

\title{Pattern formation and transition in complex networks}

\author{Dongmei Song}
\affiliation{School of Physics and Information Technology, Shaanxi Normal University, Xi'an 710062, China}

\author{Yafeng Wang}
\affiliation{School of Physics and Information Technology, Shaanxi Normal University, Xi'an 710062, China}

\author{Xiang Gao}
\affiliation{School of Physics and Information Technology, Shaanxi Normal University, Xi'an 710062, China}

\author{Shi-Xian Qu}
\affiliation{School of Physics and Information Technology, Shaanxi Normal University, Xi'an 710062, China}

\author{Ying-Cheng Lai}
\affiliation{School of Electrical, Computer, and Energy Engineering, Arizona State University, Tempe, Arizona 85287, USA}
\affiliation{Department of Physics, Arizona State University, Tempe, Arizona 85287, USA}

\author{Xingang Wang}
\email[Email address: ]{wangxg@snnu.edu.cn}
\affiliation{School of Physics and Information Technology, Shaanxi Normal University, Xi'an 710062, China}

\begin{abstract}

Dynamical patterns in complex networks of coupled oscillators are both of
theoretical and practical interest, yet to fully reveal and understand the 
interplay between pattern emergence and network structure remains to 
be an outstanding problem. A fundamental issue is the effect of network 
structure on the stability of the patterns. We address this issue by using 
the setting where random links are systematically added to a regular lattice 
and focusing on the dynamical evolution of spiral wave patterns. As the 
network structure deviates more from the regular topology (so that it 
becomes increasingly more complex), the original stable spiral
wave pattern can disappear and a different type of pattern can emerge.  
Our main findings are the following. (1) Short-distance links added to a 
small region containing the spiral tip can have a more significant effect 
on the wave pattern than long-distance connections. (2) As more random 
links are introduced into the network, distinct pattern transitions can occur, 
which include the transition of spiral wave to global synchronization, to a 
chimera-like state, and then to a pinned spiral wave. (3) Around the transitions 
the network dynamics is highly sensitive to small variations in the 
network structure in the sense that the addition of even a single 
link can change the pattern from one type to another. These findings
provide insights into the pattern dynamics in complex networks, a problem
that is relevant to many physical, chemical, and biological systems. 

\end{abstract}
\date{\today}

\maketitle

\section{Introduction} \label{sec:intro}

Pattern formation is ubiquitous in spatiotemporal dynamical systems
in nature~\cite{CH:1993,CG:book,Hoyle:book} ranging from granular 
materials~\cite{AT:2006} to ecosystems~\cite{RvdK:2008} and 
plants~\cite{WS:2004}. Complex dynamical networks, such as 
networks of coupled oscillators, are naturally spatiotemporal
systems. The past two decades have witnessed a rapid growth of research 
on various types of dynamical processes in complex 
networks~\cite{Newman:book,BBV:book}, which include 
synchronization~\cite{LHCS:2000,GH:2000,JJ:2001,BP:2002,BBH:2004,CHAHB:2005,ZK:2006a,HPLYY:2006,WHLL:2007,
GWLL:2008}, virus spreading~\cite{PSV:2001,EK:2002,WMMD:2005,CBBV:2006,
GJMP:2008,WGHB:2009,MA:2010,BV:2011}, traffic flow~\cite{ADG:2001,
EGM:2004,ZLPY:2005,WWYXZ:2006,MGLM:2008,YWXLW:2011,MB:2012},
and cascading failures~\cite{ML:2002,ZPL:2004,GC:2007,Gleeson:2008,
HLC:2008,YWLC:2009,BPPSH:2010,PBH:2010,PBH:2011,LWLW:2012}.
In these studies, a primary issue was to address the interplay between
the dynamical processes and the network structure.
An interesting problem thus concerns pattern formation in 
complex networks, where a basic question is how the network structure 
affects the dynamical patterns. To our knowledge, in spite of the vast 
literature on dynamics in complex networks, a systematic study of the 
interplay between pattern formation and network structure is lacking.
The purpose of this paper is to fill this knowledge gap by presenting
results on pattern emergence, evolution, and transitions on networks
undergoing systematic random structural perturbations.    

To probe into the interplay between network structure and dynamical
patterns in a concrete manner, we focus on coupled oscillator
networks. An interesting phenomenon in such dynamical networks is that,
under certain conditions, the oscillators can be self-organized to form 
spatial patterns~\cite{CG:book}. As the formation of the patterns relies 
heavily on the symmetry of the coupling structure of the network, 
an intuitive thinking would suggest that it is unlikely for complex 
networks to generate dynamical patterns~\cite{PHL:2007,WGLLL:2009}. 
Yet, in natural and man-made systems, there are situations where 
well-organized patterns can form on complex networks, such as the firing 
patterns in the human brain~\cite{Basar:book}. A paradox was then how 
spatially ordered patterns can emerge from random or disordered coupling 
structures associated with a complex network. There have been previous
efforts devoted to resolving this paradox. For example, pattern formation in 
complex network of coupled activators and inhibitors was studied, where 
Turing-like patterns were observed~\cite{NM:2010}. Complex networks of 
coupled excitable nodes were also studied~\cite{QHHL:2010} with respect 
to pattern formation in which the technique of dominant phase advanced 
driving was introduced, leading to the discovery of target-like wave patterns. 
Desynchronization patterns in complex network of coupled chaotic oscillators 
were subsequently studied~\cite{FZZW:2012}, where it was found that 
reordering network nodes according to the eigenvector of the unstable mode 
can be effective at identifying the stable synchronous pattern from an 
asynchronous state. Recently, computational graph algorithms were introduced
into the field of network synchronization to study synchronous patterns 
in large-size complex networks~\cite{PSHMR:2014}, where the important 
role of network symmetry in pattern formation was elucidated. In spite
of the existing works, many questions concerning pattern formation and 
transition in complex networks remain, especially with respect to 
relatively more sophisticated patterns with complex spatial structures 
such as spiral waves~\cite{HKH:2014}. 

The starting point of our study is then spiral waves in coupled oscillator 
networks, which are patterns observed ubiquitously in physical, chemical and 
biological systems~\cite{CG:book}. Different from other types of patterns 
such as Turing patterns, the stability of a spiral wave depends crucially on 
the motion of the spiral tip~\cite{Barkley:1994,GLO:2002},
leading to the specially designed methods for analyzing and controlling spiral waves~\cite{CG:book}. 
In most previous works, spiral waves are studied for networks with a regular 
spatial structure, such as a periodic lattice. Nevertheless, there were 
works on spiral waves in systems with an irregular spatial 
structure~\cite{Panfilov:2002,HHZRG:2002,WLJO:2004,SSK:2007}. For example,
the formation of spiral waves in a medium possessing random  
(small-world) connections was studied~\cite{HHZRG:2002} with the finding
that, while the structural irregularity 
is detrimental to forming and sustaining a spiral wave, a small number of 
random links can counterintuitively enhance the wave stability.
Another work~\cite{WLJO:2004} revealed that, random connections 
added locally to a regular medium can cause the meandering motion of the 
spiral tip to approach a fixed point. It was later found~\cite{SSK:2007} 
that random connections introduced globally into a regular medium can lead
to rich behaviors in the transition of the system dynamics among global 
synchronization state, steady state, and multiple spirals. These previous
works indicated that the network structure can have a significant effect
on pattern formation and transition, calling a systematic study of this issue. 

To facilitate computation, we use coupled map 
lattices (CML)~\cite{Kaneko:book} to investigate the dynamical responses of 
spiral waves to structural perturbations. Historically, CMLs were used 
to understand patterns including spiral waves in complex media such as
granular materials~\cite{VO:1998,VO:2001,HL:2008}. In our work, staring from 
a two-dimensional regular lattice capable of generating stable spiral waves, 
we systematically introduce random links into the network and study 
the transitions in the pattern dynamics as the network structure becomes 
increasingly random (complex). We uncover dynamical patterns and richer 
bifurcations that were not seen in previous 
works~\cite{HHZRG:2002,WLJO:2004,SSK:2007}, such as chimera-like states 
where two synchronization clusters coexist with many asynchronous 
oscillators and the pinned multi-armed spirals in which the arms of the 
spiral are pinned to a square-shape boundary. An intriguing finding is 
that, in the region where pattern transitions occur,
the dynamics is extremely sensitive to small changes
in the network structure. These findings shed new lights on pattern
behaviors in complex networks, which may lead to effective methods of 
pattern control.

In Sec.~\ref{sec:model}, we introduce our CML system that is capable of
generating spiral waves, and describe our strategy to supply random 
links. In Sec.~\ref{sec:transition}, we investigate pattern  
transitions as induced by a systematic change in the network structure, 
and present evidence of two types of patterns that have not been 
uncovered previously. In Sec.~\ref{sec:sensitivity}, we demonstrate 
the sensitivity of the patterns in the transition regions to small
structural perturbations and elucidate the topological properties of the 
critical links. A discussion is presented in Sec.~\ref{sec:discussion}.

\section{Model and method} \label{sec:model}

\paragraph*{Network model.}
Our CML model with a general network structure reads~\cite{KLOP:1994,Kaneko:book}
\begin{equation} \label{eq:model}
x_{i}(n+1)=(1-\varepsilon)f[x_{i}(n)]+\frac{\varepsilon}{k_i}
\sum_{j=1}^N a_{ij} f[x_{j}(n)],
\end{equation}
with $i, j=1,\ldots,N$ are the nodal indices, $N^2$ is the system size, 
$x_{i}$ is the state of the $i$th node at time $n$, and $\varepsilon$ 
is the coupling parameter. The isolated dynamics of node $i$ is governed 
by the nonlinear equation $x_i(n+1)=f[x_i(n)]$. The coupling structure of
the system is characterized by the adjacency matrix $A$ whose elements
are given as: $a_{ij}=1$ if maps $i$ and $j$ are directly connected 
and $a_{ij}=0$ otherwise. The degree of node $i$, the total number of
links attached to it, is $k_i=\sum_j a_{ij}$. 

\begin{figure}[bpt]
\centering  
\includegraphics[width=\linewidth]{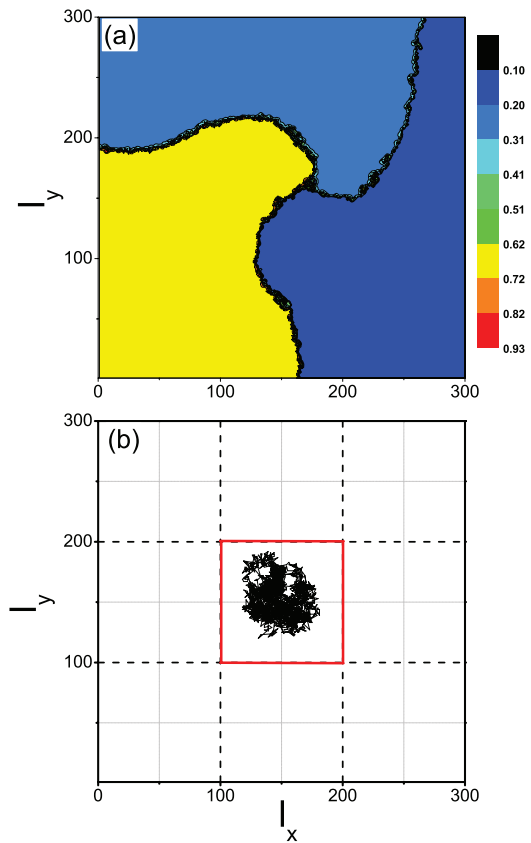}
\caption{(Color online) {\bf An example of spiral wave in the regular 
lattice}. For $\varepsilon=0.166$, a three-armed spiral 
wave generated by the system [Eqs.~(\ref{eq:model}) and (\ref{eq:fx})]: 
(a) a snapshot of the system dynamics taken at $n=3\times 10^3$, (b) random 
motion of the spiral tip inside the central area marked by the (red) square.}
\label{fig:3arm}
\end{figure}

Initially, the network is a two-dimensional regular lattice, where each 
interior node is coupled to its four nearest neighbors. We assume the free 
boundary condition and fix the system size to be $300\times 300$. The spatial
location of a node in the network is $(l_x,l_y)$, where $1\le l_x,l_y\le N$. 
For nodal dynamics, we adopt the piecewise linear map:
\begin{equation} \label{eq:fx}
f(x)=\left\{
\begin{array}{ll}
ax, & x<x_{\rm g},\\
b, & x\geq x_{\rm g},
\end{array}
\right.
\end{equation}
where $x\in(0,1)$, $x_{\rm g}=1/a$, $a$ and $b$ are independent parameters. 
Equation~(\ref{eq:fx}) is the discrete version of the differential Chay 
model used widely in computational neuroscience, and is capable of generating 
the similar bifurcation scenario of inter-spike interval (ISI) observed 
from experiments~\cite{MLWYGQR:2010}. To be concrete, we fix 
$(a,b)=(2.5,0.1)$, for which an isolated node possesses a super-stable 
period-three orbit~\cite{QWH:1998,YWQ:2015}: $x_1^* =b \equiv A$, 
$x_2^* = ab \equiv B$, and $x_3^* = a^2b \equiv C$. 

\paragraph*{Generation of spiral waves.}
For a CML system, spiral waves can be stimulated through special initial 
conditions~\cite{KLOP:1994}. For example, we can choose the initial 
states of the nodes randomly within a narrow strip in the lattice, say 
$l_x\in[130,170]$, with values uniformly distributed in the unit interval,
while nodes outside the strip are set to have the initial value zero.
For $\varepsilon=0.166$, after a transient period of $n=2.5\times 10^3$ 
iterations, a stable three-armed spiral pattern is generated, as shown
in Fig.~\ref{fig:3arm}(a), which is a snapshot of the system state. As the system 
evolves, the spiral arms rotate in a synchronous fashion and propagate 
outward from the tip. This feature of wave propagation is similar to 
that of spiral waves observed in other contexts, e.g., excitable 
media~\cite{CG:book}. A close examination of the motion of the spiral 
tip reveals a difference: in an excitable medium the tip trajectory is
often regular~\cite{LOPS:1996}, but in our system it moves randomly inside 
the central region $100<l_x,l_y<200$, as shown in Fig.~\ref{fig:3arm}(b). 
In addition, in regions separated by the spiral arms, the nodes are 
synchronized into three distinct clusters, with nodes in each cluster 
being synchronized to the trajectory of a periodic point of the period-three 
orbit. The synchronous clusters have approximately the same size. The 
spiral arms themselves comprise asynchronous nodes, which constitute 
the cluster boundaries. We use the spiral pattern in Fig.~\ref{fig:3arm}(a) 
as the initial state and investigate the effect of randomly added links
into the regular lattice.

\begin{figure}[tpb]
\centering
\includegraphics[width=\linewidth]{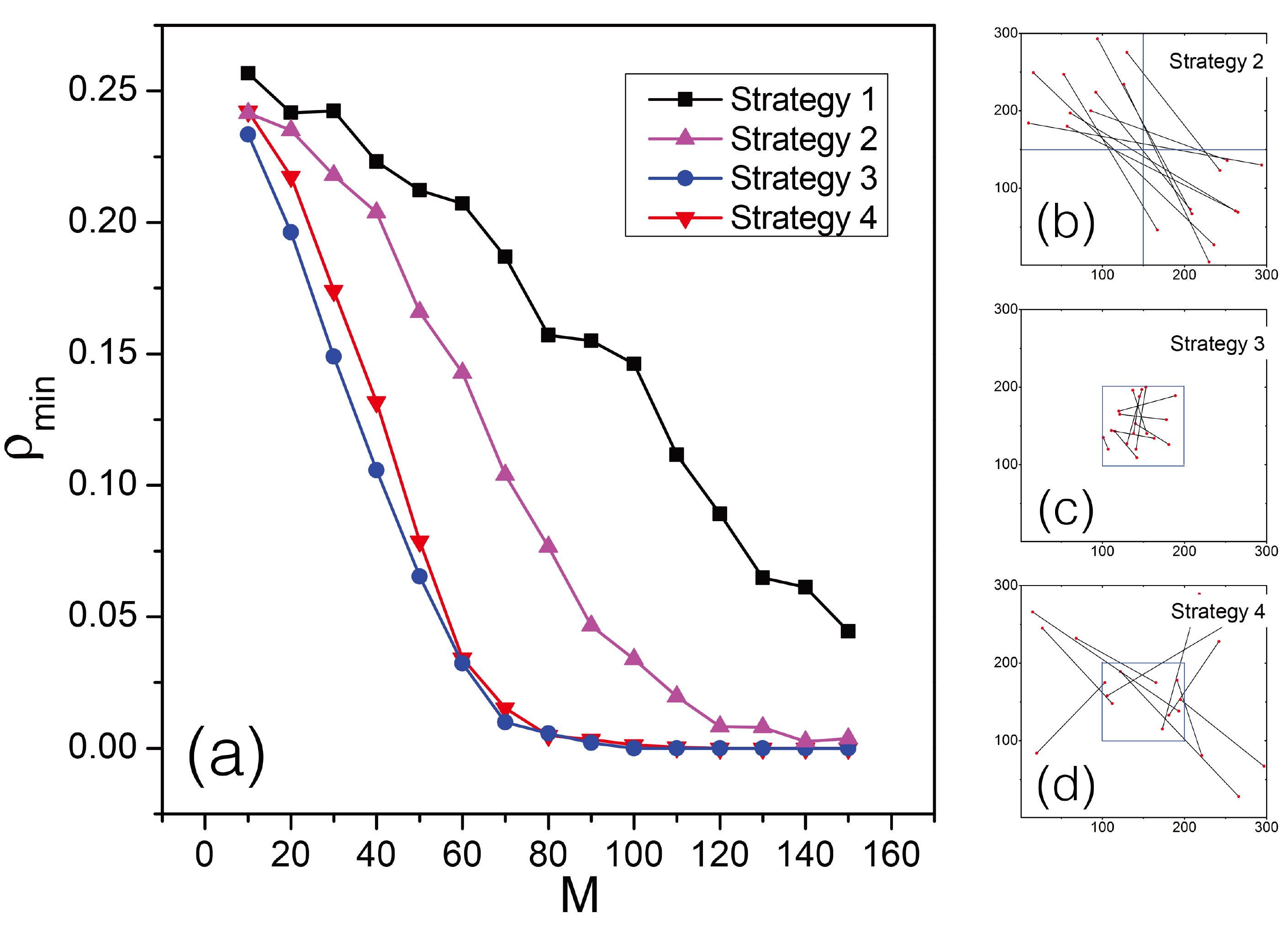}
\caption{(Color online) {\bf Sustainability of spiral wave subject to 
different types of network structural perturbations}. (a) The variation in the 
normalized size of the smallest synchronous cluster, $\rho_{min}$, with 
$M$, the number of random links. Strategy 1: adding random links over
the entire network (filled squares); Strategy 2: introducing long-distance 
random links (filled upper triangles); Strategy 3: supplying random links 
only in the central area (filled circles); Strategy 4: distributing 
random links between the central and peripheral regions (filled downward 
triangles). (b-d) Schematic illustrations of strategies 2, 3 and 4. Results 
in (a) are averaged over $200$ network realizations.} 
\label{fig:rho}
\end{figure}

\paragraph*{Effect of adding random links on spiral wave patterns.} 
As the dynamics of the spiral wave is slaved to the tip, applying random
structural perturbations to the tip region can induce characteristic
changes in the wave pattern~\cite{Barkley:1994,GLO:2002}. 
To gain insights, we first supply links between randomly chosen, unconnected 
pairs of nodes over the entire network~\cite{HHZRG:2002,WLJO:2004,SSK:2007}.
The total number of such links is $M$, with the same coupling function
and parameter as the regular links. For a fixed value of $M$, the network
is initialized with the spiral wave pattern in Fig.~\ref{fig:3arm}(a) and 
the network state is recorded after $5\times 10^3$ iterations. To characterize
the deterioration of the perturbed spiral, we introduce the quantities 
$\rho_{max}=\max\{N_A,N_B,N_C\}/N^2$ and $\rho_{min}=\min\{N_A,N_B,N_C\}/N^2$, 
with $N_A$ ($N_B$, $N_C$) being the number of nodes with state $A$ (B,C), 
which represent, respectively, the normalized size of the largest and the 
smallest synchronous clusters associated with the spiral pattern. For the 
initial spiral [Fig.~\ref{fig:3arm}(a)], as the three clusters have 
approximately the same size, we have $\rho_{max}\approx\rho_{min}\approx1/3$. 
If the sizes of the clusters are different, we have $\rho_{max}>1/3$ and 
$0<\rho_{min}<1/3$, indicating a deformed but still sustained spiral. When a
cluster disappears, we have $\rho_{min}\approx 0$, but the value of 
$\rho_{max}$ may either be close to unity (if the system reaches global 
synchronization) or $1/2$ (if two equal-size clusters coexist). In this
case, the original spiral has been destroyed. A simple criterion to 
determine the destruction of the spiral wave thus is $\rho_{min}\approx 0$. 
Figure~\ref{fig:rho}(a) shows the variation of $\rho_{min}$ with 
$M$ (strategy 1). We see that, as $M$ is increased from $0$ to $150$, $\rho_{min}$ 
decreases from the value of about $0.27$ to $0.05$. Statistically, the 
spiral wave can sustain with less than $150$ random links.

In order to understand the role played by the spiral tip in the pattern 
stability, we study three alternative perturbation 
strategies~\cite{HHZRG:2002,WLJO:2004,SSK:2007}: introducing 
long-distance random links (Strategy 2), supplying random links only in 
the central area (Strategy 3), and distributing random links between the 
central and peripheral regions (Strategy 4). These three strategies are
schematically illustrated in Figs.~\ref{fig:rho}(b-d), respectively.
For strategies 2 and 3, the pairs of nodes connected by the new links are 
randomly chosen from the regions $(0,100)\times (200,300)$ and 
$(200,300)\times (0,100)$ on the lattice, and from the central area 
$(100,200)\times (100,200)$, respectively. For strategy 4, one map is 
randomly chosen from the central area while another is chosen from 
the peripheral region. The results of applying perturbation strategies 2-4 
are shown in Fig.~\ref{fig:rho}(a), where we see that, strategies 3 and 
4 are more effective at suppressing the spiral wave pattern than
strategies 1 and 2. For example, the value of $\rho_{min}$ is reduced 
to $0$ at about $M=140$ for strategy 2; while for strategies 3 and 4, this 
occurs at about $M=100$. For the rest of the paper we focus on strategy 3.

\section{Pattern transitions} \label{sec:transition}

\begin{figure}[tbp]
\centering
\includegraphics[width=\linewidth]{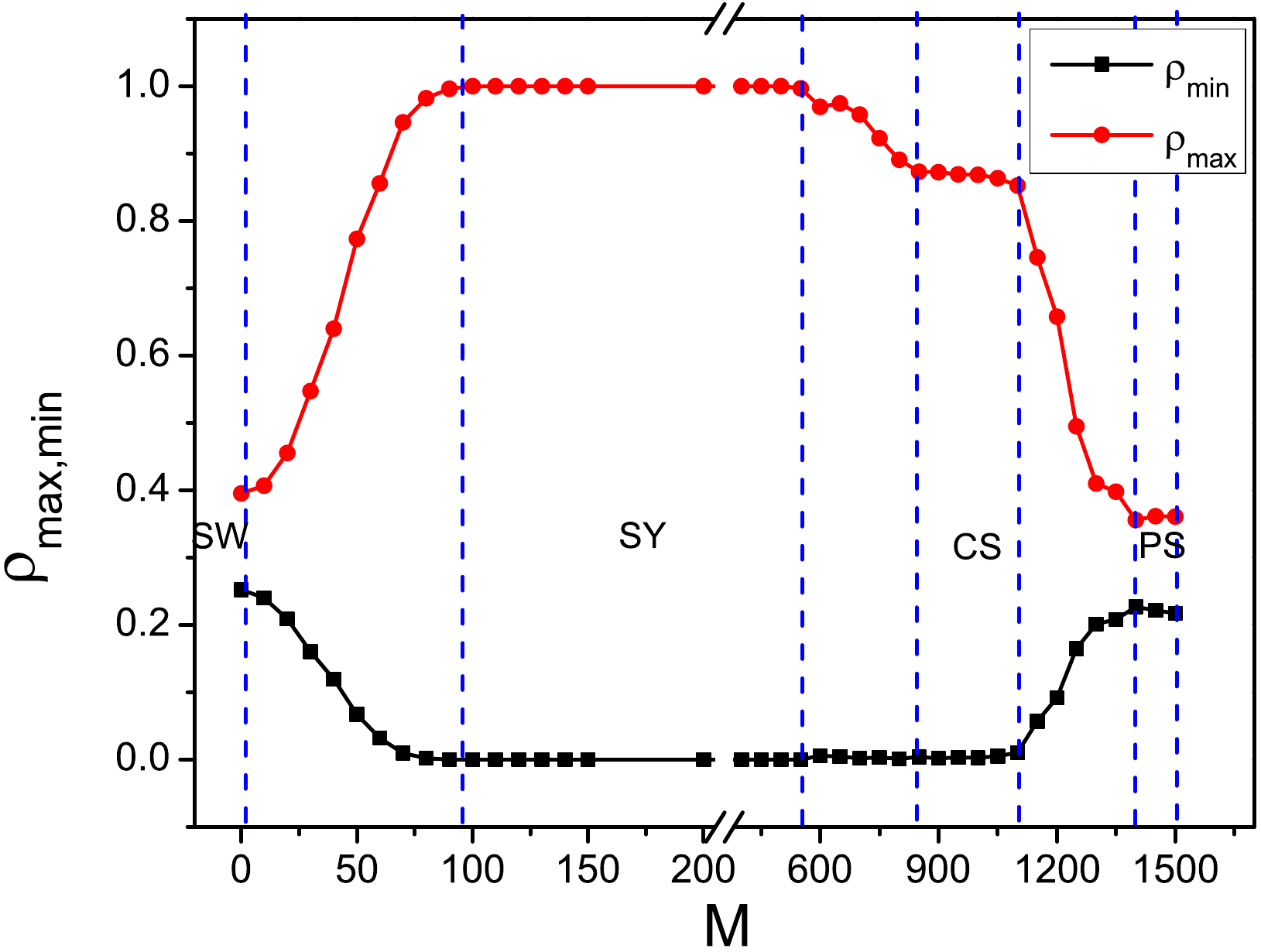}
\caption{(Color online) {\bf Pattern transitions as the network structure becomes increasingly random}. Shown are $\rho_{max}$ and $\rho_{min}$ (the 
normalized sizes of the largest and the smallest synchronous cluster, 
respectively) versus $M$, the total number of randomly added links 
according to perturbation strategy 3. Spiral wave (SW) patterns exist in 
the parameter interval $M\in [0,10]$. Global synchronization (SY) occurs
for $M\in[100,550]$. Chimera-like state (CS) arises in $M\in[850,1100]$. 
For $M > 1400$, there is pinned spiral (PS). Ensemble average of $200$ 
network realizations is used.}
\label{fig:transition}
\end{figure}

\begin{figure}[tbp]
\centering
\includegraphics[width=\linewidth]{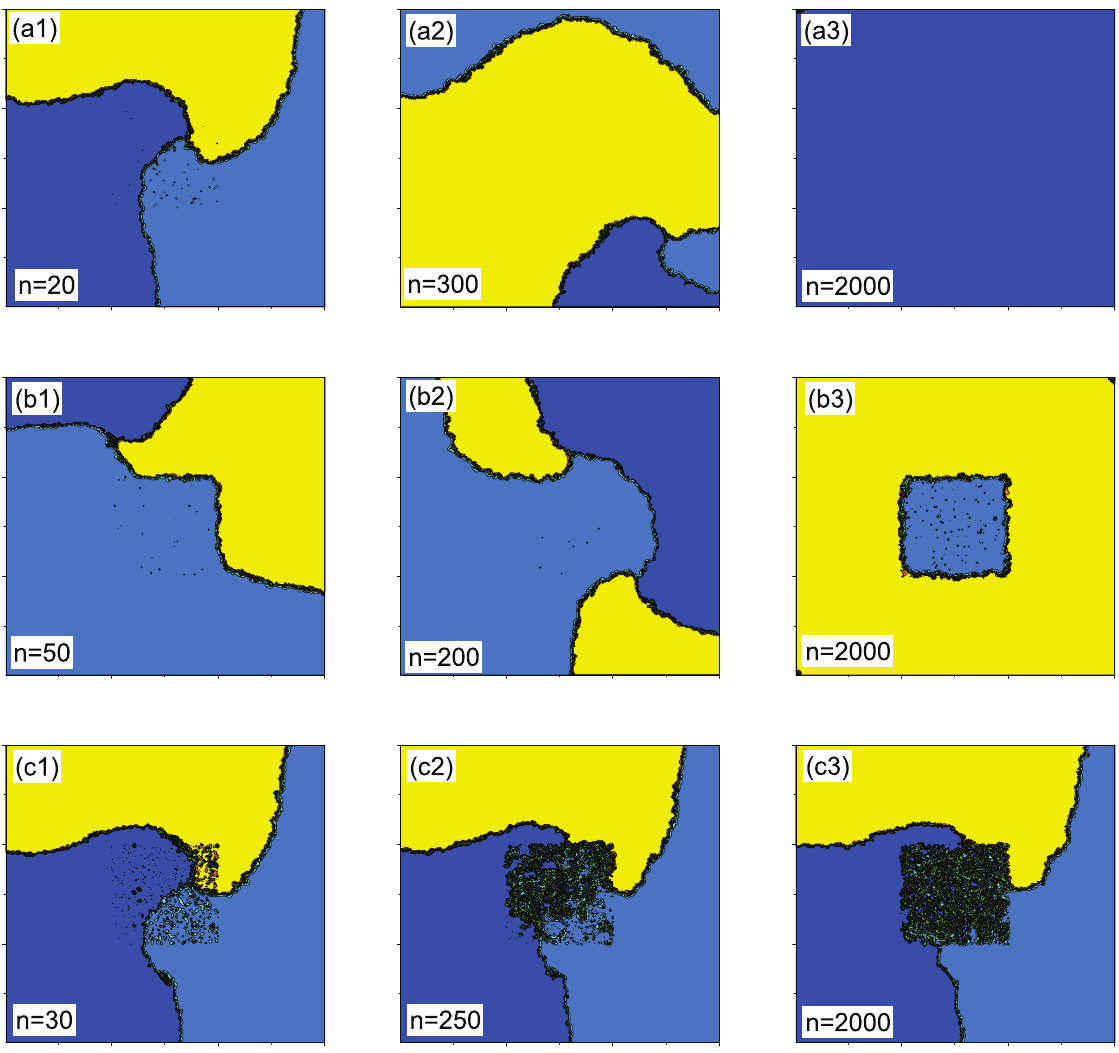}
\caption{(Color online) {\bf Characteristically distinct dynamical states
in the network system that structurally becomes increasingly random} for 
(a1-a3) $M=100$ (global synchronization) (b1-b3) $M = 1000$ (chimera-like 
state) and (c1-c3) $M = 1500$ (pinned spiral wave). The left, middle and 
right columns are snapshots of the system dynamics taken, respectively, at 
the initial, middle and end of the evolution. Black dots in the central area 
mark the ending nodes of the randomly added links.}
\label{fig:tip}
\end{figure}

To uncover and understand the pattern transitions as the network topology 
deviates from that of a regular lattice and becomes increasingly random,
we calculate the variations of $\rho_{max}$ and $\rho_{min}$ with $M$ (the 
total number of randomly added links according to perturbation strategy 3). 
As shown in Fig. \ref{fig:transition}, for a few randomly added links, say 
$M<10$, we have $\rho_{max} \approx \rho_{min} \approx 1/3$. In this region,
the network exhibits a stable spiral wave similar to that in 
Fig.~\ref{fig:3arm}(a). As $M$ is increased from $10$, the value of 
$\rho_{max}$ increases but $\rho_{min}$ decreases. For $M \approx 100$, we 
have $\rho_{max}=1$ and $\rho_{min}=0$, signifying that the network has 
reached a uniform synchronization state without any spatial pattern. 
Figure~\ref{fig:tip}(a) shows, for $M=100$, with time the spiral tip shifts 
from the central to the peripheral area [Fig.~\ref{fig:tip}(a2)]. The 
spiral tip vanishes when it reaches the lattice boundary. Subsequently, one 
of the synchronous clusters expands while the other two clusters are 
pushed toward the boundary, leading finally to the state of global 
synchronization, as shown in Fig.~\ref{fig:tip}(a3). 

Figure~\ref{fig:transition} indicates that the global synchronization state 
is stable for $M \alt 550$. As $M$ is increased further, $\rho_{max}$ 
decreases gradually but the value of $\rho_{min}$ remains about $0$. 
For $M \approx 850$, another platform emerges in the variation of 
$\rho_{max}$ with $M$, where $\rho_{max} \approx 0.9$ for 
$M\in[850,1100]$. Since $\rho_{min} \approx 0$ still holds, there is no 
spiral wave. In fact, in this region the system contains at most two 
synchronous clusters. Because the value of $\rho_{max}$ is close to unity,
most nodes in the network are synchronized into a giant cluster. To 
provide evidence for this scenario, we calculate a series of snapshots of
the system dynamics for $M = 1000$, which are shown in Fig.~\ref{fig:tip}(b).
We see that, the original spiral wave first breaks into a pair of 
anti-spiral waves [Fig.~\ref{fig:tip}(b2)] that move gradually to the 
system boundary and disappear after reaching it. In the meantime, two 
synchronous clusters emerge: a small cluster consisting of nodes in the 
central area (except for the nodes connected by the randomly added links) and
a large cluster comprising nodes in the peripheral area. There is a narrow 
boundary of asynchronous nodes separating the two clusters, as shown in 
Fig.~\ref{fig:tip}(b3). The distinct synchronous clusters 
represent effectively a chimera state observed previously in systems of non-locally coupled 
oscillators~\cite{KB:2002,AS:2004,YHLZ:2013,YHGL:2015}, which becomes unstable as $M$ 
is increased through $1100$, since $\rho_{max}$ and $\rho_{min}$ tend to
decrease and increase, respectively. For $M \agt 1400$, the values of 
$\rho_{max}$ and $\rho_{min}$ are stabilized about $1/3$ and $1/4$, 
respectively. Figure~\ref{fig:tip}(c) shows, for $M = 1400$, the typical 
states emerged during the system evolution. Due to the added random 
links, the spiral tip first drifts from the central to the peripheral 
area [Fig.~\ref{fig:tip}(c1)], but the drift stops at the boundary of 
the central area after which the tip disappears. During this time 
interval, the spiral arms are separated from each other. As 
will be demonstrated below, the three arms are attached to the boundary 
of the central area and rotate in a synchronous fashion, signifying 
the phenomenon of pinned spirals~\cite{PGFPZ:2012}. As $M$ is 
increased further, the pinned spiral state can be maintained (even
for $M=1\times 10^4$).

\begin{figure}[tbp]
\centering
\includegraphics[width=\linewidth]{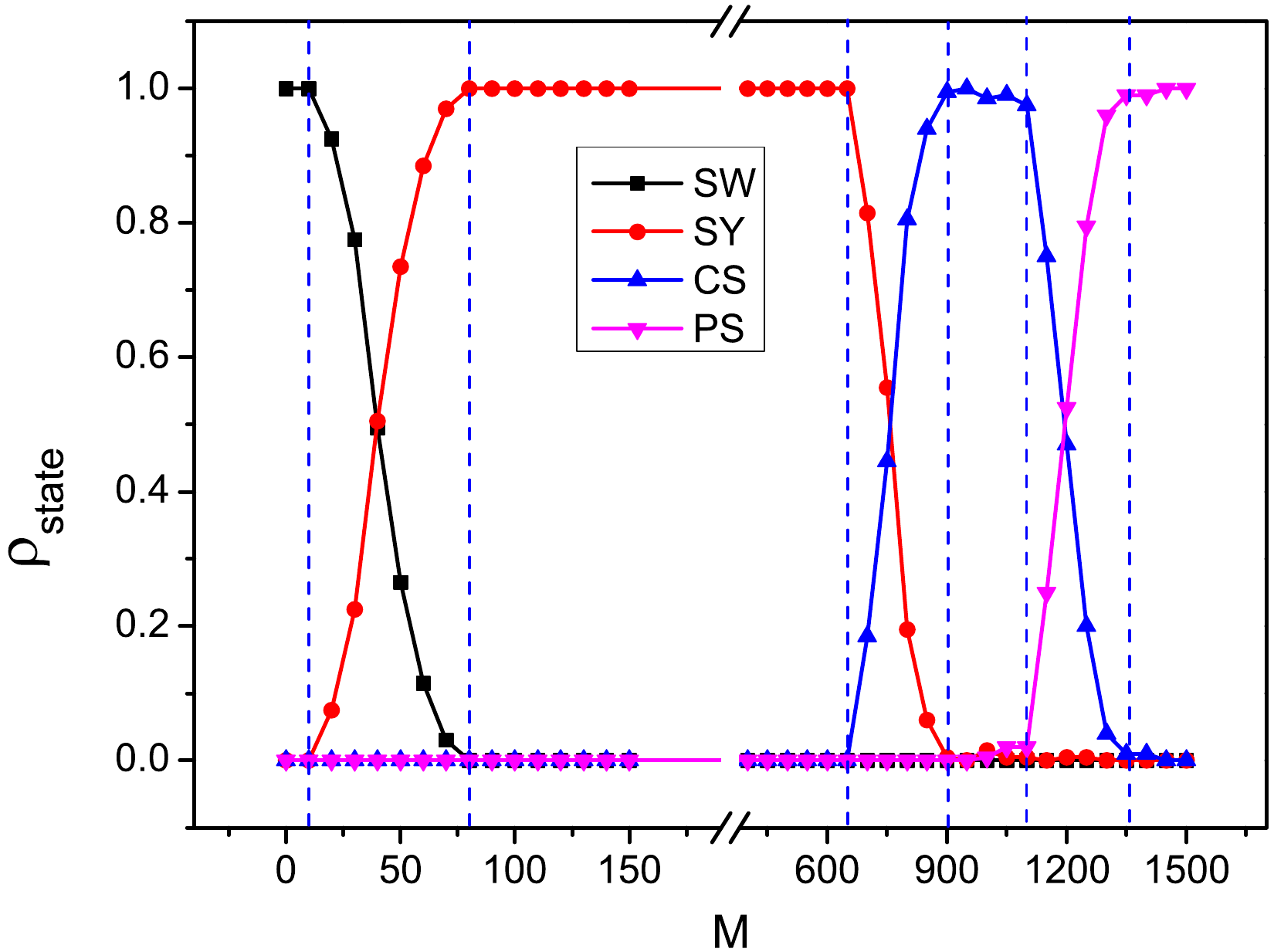}
\caption{(Color online) {\bf Continuous nature of pattern transition}.
Shown is $p_{state}$ versus $M$, which denotes the probability of 
generating a specific pattern for $M$ randomly added links. The terms SW, SY, CS and PS stand for, 
respectively, spiral wave, global synchronization, chimera-like state, 
and pinned spiral wave. Results are averaged over $200$ network realizations.}
\label{fig:probability}
\end{figure}

Figures~\ref{fig:transition} and \ref{fig:tip} suggest the following 
transition scenario as more random links are added to the network: spiral 
wave $\rightarrow$ global synchronization $\rightarrow$ chimera-like 
state $\rightarrow$ pinned spiral wave, where each state exists in a finite
parameter region. Within each region, the values of $\rho_{max}$ and 
$\rho_{min}$ are hardly changed, suggesting that the respective patterns are 
stable to random structural perturbations. A transition region is associated
with dramatic changes in the value of $\rho_{max}$ or $\rho_{min}$, 
in which one type of pattern is destroyed and a new type is born.  
To better characterize the transition regions, we calculate the 
probability of certain pattern, $p_{state}$, as a function of $M$. The 
results are shown in Fig.~\ref{fig:probability}. Comparing 
Figs.~\ref{fig:transition} and \ref{fig:probability}, we see that two different patterns coexist in each transition region. Taking $M=30$ 
as an example, we see that, over $200$ independent network realizations, 
about $80\%$ of these lead to a spiral wave ($p_{sp}\approx 0.8$) whereas 
the remaining cases correspond to the global synchronization state 
($p_{gs}\approx 0.2$). In the transition region between spiral wave and
global synchronization ($10 < M < 100$), $p_{sp}$ deceases from unity to
zero, which is accompanied by an increase in $p_{gs}$ in the opposite 
direction. This feature of gradual and continuous transition appears also
in other transition regions, as shown in Fig.~\ref{fig:probability}.   

\section{Pattern sensitivity in the transition regions} \label{sec:sensitivity}

\begin{figure}[tbp]
\centering
\includegraphics[width=\linewidth]{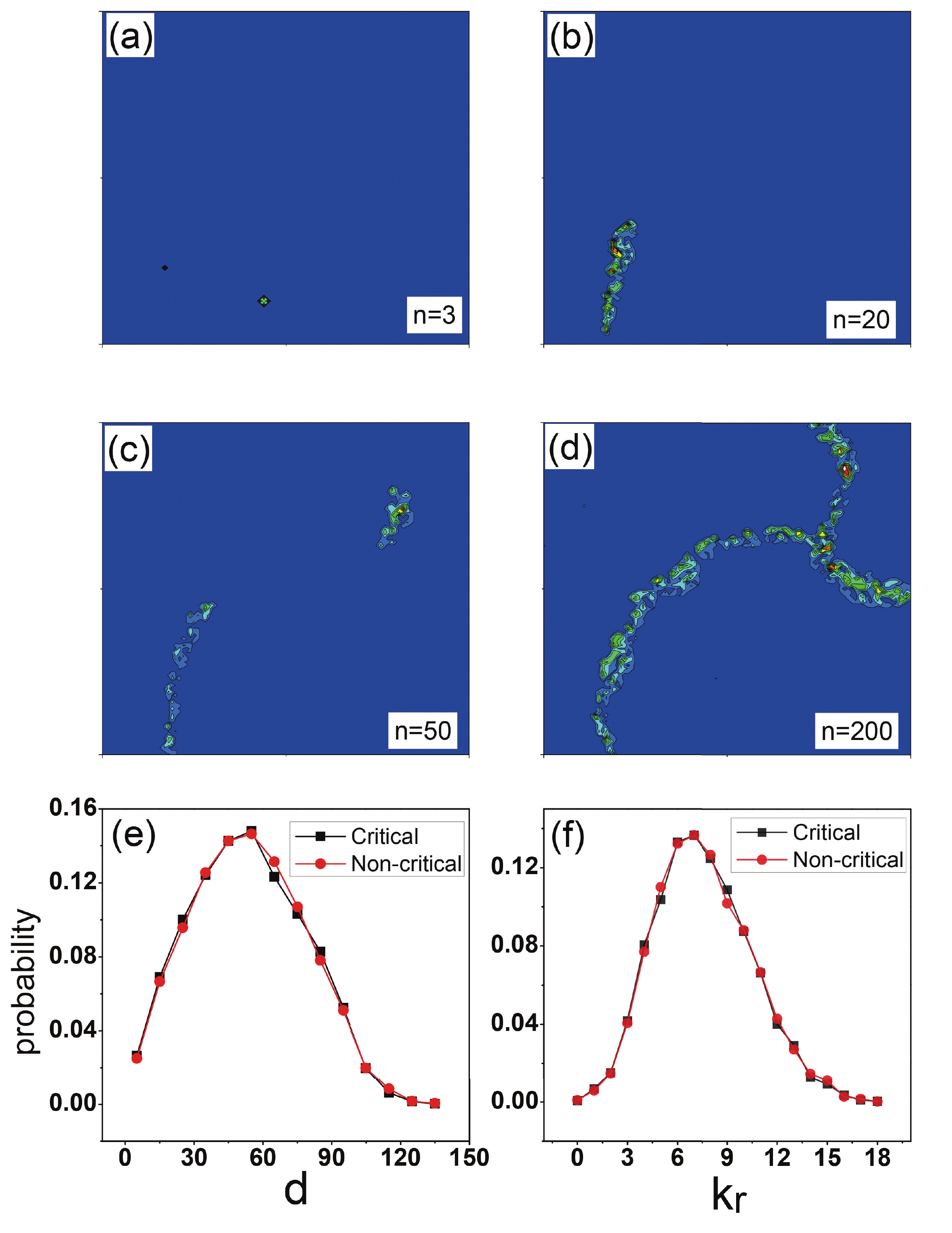}
\caption{(Color online) {\bf Pattern sensitivity in the transition regions}. 
(a-d) Time evolution of the difference $\delta \mathbf{X}$ between the 
spiral wave ($M=49$) and the global synchronization state ($M=50$). 
(e) Normalized distance distribution of the critical (filled squares) 
and non-critical links (filled circles). (f) Normalized distribution 
of the local connectivity of the critical (filled squares) and 
non-critical links (filled circles). Results in (e) and (f) are averaged 
over $500$ realizations of the $50$th link.}
\label{fig:evolution}
\end{figure}

In the transition regions the system dynamics is sensitive to random
structural perturbations in the sense that a single added random link is
able to switch the system dynamics from one pattern to another. For example, 
for a network with $M=49$ random links, the stable state of the system 
is a spiral wave pattern. With a new random link, the state of global
synchronization emerges and becomes stable, as exemplified in 
Figs.~\ref{fig:evolution}(a-d), where the time evolutions of different pattern 
states, $\delta \mathbf{X}=\mathbf{X}_{M=49}-\mathbf{X}_{M=50}$, are 
shown, with $\mathbf{X}_{M}=\{x_i\}$ denoting the pattern state for the
network with $M$ random links. Initially [Fig.~\ref{fig:evolution}(a)], 
except for the pair of nodes connected by the $50$th link, the two 
patterns are essentially identical as the two networks (one with $49$ and
another with $50$ random links) start from the same initial condition 
[the spiral wave in Fig.~\ref{fig:3arm}(a)]. Then, as the two system evolves, the difference at one of the ending node of the $50$th link is disappeared, whereas the difference of the other ending node is kept and gradually propagated to its neighboring region [Fig.~\ref{fig:evolution}(b)]. At a later time the difference switches
to a remote ending node of a random link in the lattice 
[Fig.~\ref{fig:evolution}(c)]. This propagation and switching process 
occurs repeatedly in the central area, resulting in the formation of a 
three-armed spiral, as shown in Fig.~\ref{fig:evolution}(d). Finally, 
a spiral wave similar to that in Fig.~\ref{fig:3arm} is generated. Similar behaviors arise in other transitional
regions [$M\in(550,850)$ and $M\in(1100,1400)$].    

Do the critical connections possess any special topological property? 
To address this question, in the first transition region, we add the 
$50$th link in the central area randomly. If the system evolves 
finally into the state of global synchronization, we mark this link
as critical and record the locations of the ending nodes, denoted by
$(l_x,l_y)$ and $(l'_x,l'_y)$. For comparison, we also record the ending 
nodes of non-critical links that do not lead to the destruction of the
spiral wave. We first examine the Euclidean distances of the critical 
links, defined as $d=[(l_x-l'_x)^2+(l_y-l'_y)^2]^{1/2}$. Previous 
studies~\cite{Qian:2012,XHLDL:2013} revealed that long distance links have a
more significant effect on the network dynamics than short range
links. A higher probability for a critical link to be long ranged can
then be intuitively expected. Figure~\ref{fig:evolution}(e) shows the 
normalized distribution of the distances of the critical links. 
Surprisingly, the distribution is unimodal with the maximum probability 
occurring at about $d=50$. For comparison, the distance distribution 
of the non-critical is also shown, where we see that the two 
distributions are nearly identical. This analysis indicates that 
the critical links are uncorrelated with the distance.

The local connectivity of the ending nodes associated with the critical 
links represents another topological feature. To examine it, for each 
ending node, we count the number $n_r$ of nodes of degree larger than 
$4$ within the distance $d_r=10$ from it, for the reason that there is at least one
random link attached to such a node. If the critical links were attached 
to the existing nodes following the preferential attachment rule, such
links would be more likely to lie in regions containing large degree nodes. 
In this case, the distribution of $n_r$ should exhibit a heavy tail. 
A representative distribution of $k_r$, the sum of the critical links of 
the ending nodes, is shown in Fig.~\ref{fig:evolution}(f), which exhibits
a unimodal feature. For comparison, the $k_r$ distribution of the non-critical 
links is also shown, which is indistinguishable from that associated with 
critical links. We find that the distance $d_r$ has no effect on the unimodal
feature. The results thus suggest that local connectivity is uncorrelated
with the critical links. 

Examination of additional topological properties~\cite{Newman:book} such 
as the degree assortativity, network modularity, and average network 
diameter revealed no clear difference between the critical and 
non-critical links. Topological analyses have also been carried out 
for other transition regions, with the results essentially the same. Our
conclusion is that there is little difference between the critical and
non-critical links.   

\section{Discussion} \label{sec:discussion}

In a regular lattice of coupled nonlinear oscillators, a typical (or default)
pattern is spiral wave. A sufficient number of random links will destroy
the spiral wave, but how? This paper addresses this question through a 
detailed computational study of the effect of random links on a dynamical
pattern as their number is systematically increased. From the point of view of 
network structure, adding random links to a regular network makes it 
complex, thus our work effectively addresses the general problem of 
pattern formation and transition in complex networks. We find, as the 
number of randomly added links is increased, the underlying networked 
system can exhibit distinct types of dynamical patterns and rich transitions
among them. Our study has revealed two types of patterns in complex 
networks, which to our knowledge have not been reported previously: 
a chimera-like pattern and pinned multi-armed spiral waves. We also 
find that a transition between two distinct types of patterns can be 
triggered through only a single, critical random link. The topological
properties of the set of critical links are found to be no different 
from those of the non-critical links. With respect to spiral waves, our 
study reveals that random links added into the region of the spiral tip 
can have a devastating effect on the pattern, a result that is consistent 
with those from previous works~\cite{HHZRG:2002,WLJO:2004,SSK:2007}. 

A key difference from previous studies is that in our system 
the motion of the spiral tip is random in a limited area (the central area), 
which leads to the sensitivity of the system pattern in the transition regions. This has  
a consequence. In particular, a previous result in the study of 
synchronization transitions in complex networks was that, as the network 
becomes increasingly complex, an order parameter characterizing the 
degree of network synchronization can increase 
continuously~\cite{ADKMZ:2008}. As demonstrated in this work,
for spiral wave patterns a properly defined order parameter would 
exhibit a non-monotonous behavior because, as the number of random links is 
increased, the state of the network can become perfectly ordered (global 
synchronization) and then become less ordered with a chimera-like state 
and a pinned multi-armed spiral. This peculiar phenomenon may be specific
to spiral-wave patterns where we supply random links only to the tip region. 
Indeed, when random links are added to the network on 
a global scale (as with the Newman-Watts small-world network 
model~\cite{NW:1999}), we find a monotonic change in the degree of
the order of the system state. Nonetheless, our study sheds new lights on 
the pattern dynamics in complex networks and our results provide insights
into the issue of pattern control on networks.

\section*{acknowledgements}

We thank Prof. H.~Zhang for helpful discussions. This work was supported 
by the National Natural Science Foundation of China under the Grant 
No.~11375109 and by the Fundamental Research Funds for the Central 
Universities under the Grant No.~GK201601001. 
YCL would like to acknowledge support from the Vannevar Bush
Faculty Fellowship program sponsored by the Basic Research Office of
the Assistant Secretary of Defense for Research and Engineering and
funded by the Office of Naval Research through Grant No.~N00014-16-1-2828.


%
\end{document}